# Coherent Erbium Spin Defects in Colloidal Nanocrystal Hosts

Joeson Wong[1,2,3,*], Mykyta Onizhuk[4], Jonah Nagura[4], Arashdeep S. Thind[5], Jasleen K. Bindra[6], Christina Wicker[4], Gregory D. Grant[3,4], Yuxuan Zhang[4], Jens Niklas[6], Oleg G. Poluektov[6], Robert F. Klie[5], Jiefei Zhang[3,7], Giulia Galli[2,3,4,7], F. Joseph Heremans[3,4,7], David D. Awschalom[3,4,7], A. Paul Alivisatos[1,2,4,*]

[1]James Franck Institute, University of Chicago, Chicago, Illinois 60637, United States

[2]Department of Chemistry, University of Chicago, Chicago, Illinois 60637, United States

[3]Materials Science Division, Argonne National Laboratory, Lemont, Illinois 60439, United States

[4]Pritzker School of Molecular Engineering, University of Chicago, Chicago, Illinois 60637, United States

[5]Department of Physics, University of Illinois Chicago, Chicago, Illinois 60607, United States

[6]Chemical Sciences and Engineering Division, Argonne National Laboratory, Lemont, Illinois 60439, United States

[7]Center for Molecular Engineering, Argonne National Laboratory, Lemont, Illinois 60439, United States

*Corresponding Authors: joeson@uchicago.edu, paul.alivisatos@uchicago.edu

## Abstract

We demonstrate nearly a microsecond of spin coherence in $Er^{3+}$ ions doped in cerium dioxide nanocrystal hosts, despite a large gyromagnetic ratio and nanometric proximity of the spin defect to the nanocrystal surface. The long spin coherence is enabled by reducing the dopant density below the instantaneous diffusion limit in a nuclear spin-free host material, reaching the limit of a single erbium spin defect per nanocrystal. We observe a large Orbach energy in a highly symmetric cubic site, further protecting the coherence in a qubit that would otherwise rapidly decohere. Spatially correlated electron spectroscopy measurements reveal the presence of $Ce^{3+}$ at the nanocrystal surface that likely acts as extraneous paramagnetic spin noise. Even with these factors, defect-embedded nanocrystal hosts show tremendous promise for quantum sensing and quantum communication applications, with multiple avenues, including core-shell fabrication, redox tuning of oxygen vacancies, and organic surfactant modification, available to further enhance their spin coherence and functionality in the future.

## Introduction

Quantum information technology promises to revolutionize sensing[1,2], communications[3,4], and computing[5,6] by using the intrinsic quantum mechanical properties of entanglement and superposition to obtain a quantum advantage. Developing quantum hardware that is both robust against decoherence and can scale to practical system sizes is a grand challenge for quantum information technology today[7,8]. New materials platforms present an exciting opportunity to address these challenges related to robustness, scaling, and integrability by providing additional functionality not seen in other systems.

Optically active spin defects in crystalline host materials are an attractive qubit representation given their natural transduction between spin and optical degrees of freedom, which act as stationary and flying qubits, respectively[9]. While many of these spin defects are deeply embedded in a solid-state host (i.e., away from interfaces and surfaces) to improve their spin coherence times, this same property makes them difficult to couple and integrate with other systems[10], such as photonic structures that could significantly improve a spin qubit's optical addressability and state read-out fidelity[11]. Meanwhile, quantum sensing requires that the sensing qubit be near the sensing target to maximize its sensitivity and spatial resolution[1,12]. Both functionalities necessitate proximity of the qubit to the host material's surface. Yet, it is well-known that solid-state surfaces often contain complex terminations, atomic reconstructions, and dangling bonds that act effectively as sources of electric and magnetic noise, deleteriously affecting optical and spin coherence[13].

Colloidal nanocrystals are a materials platform where exquisite control of surface morphology meets synthetic tunability in a system where the surface to volume ratio approaches unity. This synthetic control of surface morphology has enabled photoluminescence quantum yields approaching unity and optical coherence times approaching their transform-limit in a variety of materials systems[14–16]. Colloidal nanocrystals also provide flexibility in integrating with various platforms and structures, enabling their applications in optoelectronics[17–19], medicine and biology[20–22], nanophotonics[23,24], and catalysis[25]. Yet, investigation of optically active spin qubits in colloidal nanocrystals has so far been limited[26]. To date, only spin defects operating in the visible portion of the spectrum have been studied and their corresponding spin-photon interface remains underdeveloped[27,28]. Furthermore, a comprehensive understanding of the factors that limit the spin coherence in doped nanocrystal systems is still to be developed.

In this work, we demonstrate an ensemble-measured electron spin coherence time approaching a microsecond in telecommunication-compatible erbium spin defects embedded in cerium dioxide nanocrystal hosts. We suggest that this long spin coherence, despite the large gyromagnetic ratio of the spin defect and its proximity to the surface, is likely due to a combination of the nuclear spin free ceria host and the unique crystal field levels that separate the ground state spin level from the rest of its orbital manifold. High-resolution scanning transmission electron microscopy imaging combined with electron energy loss spectroscopy further confirms the highly crystalline nature of the host material, along with the presence of $Ce^{3+}$ on the surface of the nanocrystal. Further, we use cluster correlation expansion calculations of the qubit's density matrix to understand the unique size, position, and concentration dependence of spin defects in nanocrystal hosts to corroborate both our experimental results and to outline important principles for designing colloidal nanocrystals with long spin coherence. By simultaneously improving the spin and optical coherence via rational materials design and synthetic control, nanocrystal hosts could become a valuable platform for many future quantum information technologies.

## Results

Our system of interest consists of a colloidal nanocrystal (**Figure 1a**) with an inorganic ceria ($CeO_2$) core covered with an organic (oleate) shell. Embedded within the inorganic ceria nanocrystal matrix are isolated erbium ($Er^{3+}$) spin defects (**Figure 1b**). To synthesize our nanocrystals, we employed a two-phase solution in an autoclave heated to 180 °C, with the solutions consisting of oleic acid dissolved in toluene and rare-earth nitrates dissolved in water. We used erbium precursors with a natural abundance of isotopes (i.e., ~23% [167]Er). Incorporation of erbium ions into the ceria lattice is done via mixing of erbium and cerium nitrate precursors (see **Methods**), and the resultant molar ratio of precursor concentration to incorporated defect concentration is near unity (**Table S1**), likely due to the similar ionic radii of the respective cations (**Figure S1**)[29]. We synthesized a variety of erbium doped ceria nanocrystals with different doping levels and estimated their erbium concentration by performing both size distribution measurements in a TEM along with molar ratio measurements using either inductively coupled plasma optical emission spectroscopy (ICP-OES) or inductively coupled plasma mass spectroscopy (ICP-MS) (**Methods**). Throughout the text, $\langle N_{Er} \rangle$ refers to the estimated ensemble average number of erbium defects per ceria nanocrystal. Here, ceria also acts as a nuclear-spin free host material in its natural isotopic abundance, suggesting that long spin coherences should be achievable[30].

Isolated $Er^{3+}$ ions have 4f electrons shielded by delocalized 6s and 5p orbital electrons, decoupling them from direct interactions with the environment. The 4f electrons exhibit strong spin-orbit coupling with weak optical transitions due to their parity forbidden transitions. The ground state manifold, corresponding to $^4I_{15/2}$ in spectroscopic notation, or $Z_n$, is (2J+1) = 16-fold degenerate. Within a cubic crystal field, this 16-fold degenerate state splits into three Stark quartets $\Gamma_8$ and two different Kramers doublets $\Gamma_6$ and $\Gamma_7$[31,32]. In a similar vein, the first excited state manifold, denoted by $^4I_{13/2}$ (or $Y_n$) is 14-fold degenerate, and within a cubic crystal field, splits into two Stark quartets and three Kramers doublets. Thus, under a cubic crystal field in the absence of an applied magnetic field, we expect five distinct energy levels for both the ground state manifold $Z_n$ and the first excited state manifold $Y_n$ (**Figure 1c**). The cubic symmetry of the $Er^{3+}$ spin defect can generally be ascertained from electron paramagnetic resonance (EPR) measurements, as detailed below. Here, the lowest energy ground state $Z_1$ corresponds to one of the Kramers doublets, whose degeneracy can be further split under the application of a magnetic field. It is these two energy levels in the Kramers doublet that form the spin qubit studied here, where the $|0\rangle$ state is initialized thermally (**Figure 1d**).

To first confirm the high crystalline quality of our synthesized nanocrystals, we performed powder X-ray diffraction (PXRD) of nanocrystals drop-casted on background-free silicon holders (**Figure 2a**). The observed peaks in the X-ray diffractogram correspond well with that of bulk ceria, with an occasional low angle peak due to the ordering of the ligand shell[33]. The fitted lattice constant to a face-centered cubic lattice yielded a lattice constant of 5.418±0.002 Å. Scherrer diameters (~7 nm) were within a standard deviation of the size distribution (~1 to 1.5 nm) estimated from transmission electron microscopy (TEM) images, where mean diameters are typically 6 to 8 nm (see **Table S1** for details). Each nanocrystal therefore has roughly 3000 to 6000 cerium atoms per nanocrystal, with the precise number highly sensitive to the particular diameter of the nanocrystal. Our ceria nanocrystals were generally rhombicuboctahedron shaped, as observed via TEM (**Figure 2b**). We have further performed scanning transmission electron microscopy (STEM) to investigate the crystal structure and chemical composition of these ceria nanocrystals. Inverted annular bright field (ABF) (**Figure 2c**) and high-angle annular dark-field (HAADF) (**Figure 2d, i**, and **Figure S2**) imaging confirm the highly crystalline nature of each individual nanocrystal with clearly resolved Ce and O sub-lattice along [110] and [100] crystallographic orientations.

To probe the local chemical environment of individual nanocrystals, we have used spatially resolved electron energy loss spectroscopy (EELS) (**Figure 2d**). Multiple EEL spectra were acquired at probe positions marked in **Figure 2d, i**, whose colors correspond to the spectra shown in **Figure 2d, ii, iii, iv**. We observe marked changes in the fine structure of Ce M and O K core-loss edges at the surface (red and blue) and the bulk (cyan, black and green), which is typical of ceria nanocrystals[34–37]. The Ce $M_5$ and $M_4$ edges correspond to the transitions from an initial state of $3d^{10}4f^0$ to the final state of $3d^94f^1$. The two post edge peaks result from the transitions to 4f states in the conduction band due to hybridization between Ce 4f and O 2p states. The diminishing intensity of the post-edge peaks towards the surface (red and blue in **Figure 2d, iii**) as compared with the bulk (cyan, black and green in **Figure 2d, iii**) of the nanocrystals is caused by the reduction of $Ce^{4+}$ to $Ce^{3+}$. Also, a chemical shift of 1.5 eV to lower energy loss is observed for the Ce $M_5$ and $M_4$ edges during reduction of $Ce^{4+}$ to $Ce^{3+}$. This reduction of $Ce^{4+}$ is driven by the formation of oxygen vacancies at the surface of ceria nanocrystals[34–37]. Moreover, the reduction of $Ce^{4+}$ is also discernible in the O K edge spectra, where the absence of a pre-peak at the surface is a direct result of decreased hybridization between O 2p and Ce 4f states, arising from the formation of oxygen vacancies and simultaneous reduction of $Ce^{4+}$ to $Ce^{3+}$ at the surface of the nanocrystals. A detailed STEM-EELS dataset showing the chemical distribution of Ce and O, as well as the distribution of $Ce^{3+}$ and $Ce^{4+}$ across the surface and bulk of a ceria nanocrystal is shown in **Figure S3**. The low concentration of erbium results in a noisy Er M edge signal, which makes the analysis of Er M edge fine structure difficult. However, the EEL spectra of the Er M edge confirms erbium incorporation in the bulk and at the surface within individual ceria nanocrystals. Erbium doping was further verified via energy dispersive X-ray spectroscopy over large ensembles (**Figure S4**).

The ground state spin structure of erbium ions can be understood from EPR measurements. We performed EPR measurements for various estimated ensemble average number of erbium defects per ceria nanocrystal $\langle N_{Er} \rangle$, which is the number labelled next to each EPR spectrum in **Figure 3a**. The EPR linewidth narrows by more than an order of magnitude as the erbium dopants per nanocrystal decrease. At the lowest doping concentration of $Er^{3+}$ (i.e., $\langle N_{Er} \rangle = 0.3$), the inhomogeneously broadened linewidth is approximately 200 MHz. Further analysis of the spectrum shows clear indication of hyperfine interactions of the $Er^{3+}$ effective S = ½ electron spin with the I = 7/2 $^{167}$Er nuclei, which fits very well with numerical simulations of the spin Hamiltonian, yielding a hyperfine interaction of roughly 695 MHz (see **Methods**, **Figure S5**, and **Table S1**). However, we find large error bars on the fit for $\langle N_{Er} \rangle = 179.8$, which is associated with the larger residuals at the tails of the broad EPR spectra (**Figure S5**). Moreover, we find better fits to the experimental spectra, particularly at high erbium concentrations, by separately fitting the linewidths for the $I = 0$ and $I = 7/2$ erbium nuclei (**Figure S5**). In contrast, at low concentrations, a single broadening term for both types of nuclear spins can be used to describe the entire EPR spectrum. The crossover between the two regimes occurs when the linewidth is approximately the size of the hyperfine interaction. The g-factor of the $Er^{3+}$ ion is 6.75, very close to the theoretical value of 6.8 for $Er^{3+}$ incorporation with cubic site symmetry[31,32]. This confirms that $Er^{3+}$ substitutionally occupies a $Ce^{4+}$ site without local charge compensation[38], similar to what has been observed in bulk[32] and thin film systems[39]. Charge neutrality of the nanocrystals is likely achieved via the formation of oxygen vacancies, e.g., at the surface, as shown with STEM-EELS measurements presented earlier. However, wider magnetic field scans of the EPR spectra (**Figure S6**) do not indicate a strong signal from $Ce^{3+}$ spins despite the large quantities observed in STEM-EELS (**Figure S3**). This suggests that the environmental conditions (i.e., high vacuum, high voltage) of electron microscopy measurements are far more reducing than the environmental conditions of EPR measurements[40,41]. Therefore, the $Ce^{3+}$ concentrations measured via EELS should be used only as an upper bound when estimating their impact in EPR measurements.

The extracted EPR linewidth at low temperature as a function of erbium dopant concentration exhibits a slightly sublinear relationship (**Figure 3b**) with the number of erbium defects per nanocrystal. In bulk materials, a linear dependence with defect concentration is expected[42]. This suggests that there may be covarying factors such as size, strain, or surface effects that modify the EPR linewidth, in addition to the intrinsic dipolar contribution from $Er^{3+}$-$Er^{3+}$ electron spin interactions. Nonetheless, the EPR linewidth at the lowest erbium concentration is comparable to thin films[39,43], despite the higher erbium to cerium molar ratio, pointing to the material quality and unique form factor of the synthesized nanocrystals.

Temperature dependent measurements of the linewidths (**Figure 3c**) were performed for the lowest erbium concentration (i.e., $\langle N_{Er} \rangle = 0.3$). As expected, the line broadens as the temperature is increased. Above 40K, the signal-to-noise ratio makes it difficult to reliably estimate the linewidth. Nevertheless, the observed linewidth primarily scales exponentially with temperature, suggesting an Orbach-like phonon-mediated process of spin relaxation[8,44,45]. Examining other possible phonon processes to describe the temperature dependence observed here, such as direct and Raman mediated phonon processes, suggests their contribution is likely small (**Figure S7**). It is important to note that these phonon processes strictly speaking perturb spin relaxation and not directly the spin linewidth. Therefore, a more thorough analysis of the relative contributions of the phonon processes would require an analysis over a larger range of temperatures and direct measurements of the spin relaxation. Yet, there is precedent in extracting Orbach energies from the spin linewidth, since spin relaxation ultimately sets the maximum timescale for dephasing[44,46]. In this system, the extracted activation energy of the Arrhenius process, i.e., the Orbach energy is 8.4 $\pm$ 1.1 meV. The Orbach energy typically corresponds to the energy splitting between the ground state and a nearby orbital excited state, limiting the temperature range of qubit operation[47]. Photoluminescence excitation (PLE) spectra at 4K (**Figure S8**) indicates erbium emission from various telecommunication-compatible optical transitions. Further work is to be carried out to identify level assignments.

Pulsed EPR measurements were done to characterize the ensemble averaged erbium ion Hahn-echo sequence spin coherence times ($T_2$) and spin relaxation times ($T_1$). The Hahn-echo spin coherence times are measured for several erbium concentrations (**Figure 4b**). At the lowest concentrations, we measure a spin coherence of 776.4 $\pm$ 11.7 ns, which is large considering the gyromagnetic ratio and the proximity of the erbium ion to the surface. In fact, the magnetic noise $\delta B$ experienced by the $Er^{3+}$ spins can be estimated[48] from the gyromagnetic ratio and the measured coherence time, yielding $\delta B \sim 4.3~\mu T$. This magnetic noise is comparable to or even better than many bulk systems of rare-earth doped oxides[43,49,50]. Further, the estimated dipolar sensitivity of this spin system is on par with shallow delta-doped NV centers in diamond (**Figure S9**)[51]. Experimentally, this dipolar sensitivity is observed in the form of the envelope modulation in the spin echo intensity, which has a characteristic frequency of 4.3 MHz. Fitting of the signal with the observed envelope modulation (**Methods**) yields a frequency that agrees with the hydrogen ($^1$H) nuclear Larmor frequency at this magnetic field, showing direct evidence of sensing of the external hydrogenic nuclei on the surface of the nanocrystal (**Figure 4c**). The detected hydrogen nuclei are likely both from the oleate as well as the toluene glassy matrix the nanocrystals are embedded in, with the latter expected to contribute significantly less considering the length of the oleic acid molecule (~2 nm).

Inversion recovery echo-detected measurements were also performed to ascertain the spin relaxation dynamics (**Figure 4a**). The observed relaxation dynamics were observed to be multiexponential, and so a biexponential fit was used. Acquisition at smaller time steps improved temporal resolution and was used to improve the fit precision (**Figure S10**). The relaxation dynamics consist of a fast and slow component, with fitted values denoted in **Figure 4a**. Previous measurements in thin film samples[43] showed that the fast component of their biexponential systems could be attributed to coupling of the $Z_1$ spin states to the $Z_2$ energy levels; however this scenario may be less likely in the case of the nanocrystals, considering the Orbach energies observed. An alternative explanation is that there are multiple relaxation processes occurring, for example, due to cross-relaxation of erbium to both surface paramagnetic species (such as $Ce^{3+}$) and to other erbium spins, which, when ensemble-averaged, results in multiexponential dynamics[52,53]. Similar effects are likely at play with the spin coherence dynamics as well, as detailed by the coherence calculations described later, but detecting the multiple time scales is below the noise floor in our spin-echo measurements. Further, given the timescales observed for our spin coherence and spin relaxation dynamics, it is evident that our spin coherence at the lowest temperatures studied (~3.2K) is not limited by relaxation dynamics (i.e., $T_2 \ll 2T_1$), with potentially many avenues to improve erbium coherence times. However, we note that the observed spin relaxation and spin coherence times in this system are already comparable to and even longer than the times achieved in their thin film counterpart[43] despite the proximity to the surface, suggesting that there may be fundamental advantages and opportunities to developing long-lived qubits in nanocrystals.

To understand the factors at play that limit the spin coherence of the erbium spins in these nanocrystal hosts, we turned to numerical simulations. We performed cluster correlation expansion (CCE) calculations that determine the quantum evolution of the central erbium spin coherence embedded in a nuclear spin bath[54], to understand the effect of the hydrogen near the surface of the nanocrystals. Convergence of the coherence dynamics was ensured by examining increasingly higher orders of the CCE approximation, size of the nuclear bath, and pairwise cutoff radius (**Figure S11**). **Figure 5a** shows the dependence of the theoretical Hahn-echo signal on the position of the erbium spin in a nanosphere with a 7 nm diameter, showing the drastic impact of the atomistic location of the $Er^{3+}$ spin defect on its spin coherence. The hydrogen nuclei envelope modulation is clearly reproduced in the simulations, with the depth of the modulation being directly related to the distance to the surface. Due to the $1/r^3$ dependence of the magnetic dipolar interactions between erbium and hydrogen, as the erbium spin defect approaches the surface, the secular approximation breaks down. Non-secular flips of the hydrogen nuclear spins in the spin Hamiltonian in this regime play a very significant role, which severely reduces the erbium spin coherence time by an order of magnitude near the surface, as observed in the inset of **Figure 5a**. Considering the large surface to volume ratios of our nanocrystals and assuming an uniform substitution probability, the most likely position of an erbium spin defect in these nanocrystals is roughly within a nanometer to the surface (**Figure S12**).

We next consider the size-dependence of the nanocrystal on the achievable spin coherence, limited by the hydrogen nuclear spins (**Figure 5b**). By simulating the nanosphere diameter from 5 nm to 10 nm with an erbium spin defect located at its center, we can clearly see that the spin coherence increases with size. Importantly, our calculated values are significantly below the 47 ms predicted for bulk $CeO_2$[30], clearly showing the deleterious effects of hydrogen nuclei at the surface despite a nuclear-spin free host. Yet, spin coherence times more than 10 $\mu s$ should be readily possible for $Er^{3+}$ in a nanosized host with improved synthetic control and complexity, e.g. via shelling and ligand exchange. Developing ligands with low nuclear magnetic moments will therefore be critical to maximizing spin coherence in colloidal nanocrystal hosts. Of note is that the dependence of the coherence time with size is observed to be approximately linear, which can be understood as stemming from the competition of two scaling laws: the number of hydrogen nuclei at the surface of the nanocrystal (~$r^2$) and their dipolar coupling to $Er^{3+}$ (~$1/r^3$). Specifically, the size dependence of the nanocrystal merely comes from reducing the interactions of the central $Er^{3+}$ spin with the nuclear spin bath on the nanocrystal surface. This points to opportunities in modifying shapes of nanocrystal hosts[55], with different coherence time scaling laws possible depending on how the surface area scales in the specific geometry with dipolar coupling. Thus, improving the size and shape dispersion of nanocrystal hosts as well as the position dispersion of the embedded spin defects will be imperative for maximizing the coherence times of nanocrystal quantum hardware.

Next, we theoretically analyze the decoherence induced by instantaneous diffusion[56] of resonant erbium spin defects that are dipolar coupled, for different concentrations of erbium spins, in the absence of hydrogen nuclei-induced decoherence (**Methods**, **Figure 5c**). Here, we assume non-aggregated nanocrystals given the colloidal stability of oleate capped nanocrystals dispersed in toluene. The distribution of decoherence rates consists of two main contributions: 1. A slow decoherence rate that stems from the dipolar coupling of erbium spins belonging to different nanoparticles (i.e., interparticle induced decoherence) and 2. A fast decoherence rate that stems from erbium spins belonging to the same nanoparticle (i.e., intraparticle induced decoherence). The difference of several orders of magnitude in decoherence rates is associated with the significantly different length scales of intraparticle and interparticle interactions. The sharp features in the computed decoherence rates are due to erbium incorporating at discrete lattice sites of the ceria crystal structure, where we assume erbium can substitutionally replace cerium sites. Of note is that the effective instantaneous diffusion rate for a given molar ratio of spin defects to host atoms is significantly lower in a nanocrystal, given by its corpuscular nature compared to bulk crystals where defects are typically homogeneously distributed (**Figure S13**). This is particularly relevant since instantaneous diffusion is known to limit the spin coherence in many rare-earth doped oxide systems[43,57].

Lastly, we consider an ensemble-averaged coherence time by combining all the effects mentioned above (i.e., related to size, position, and concentration), where the main decohering interactions of a central Er spin is with other Er spins (at high concentration) and the hydrogen nuclear spins (at low concentration) (**Figure 5d**). Specifically, we simulate the experimental coherence time by averaging over the size distribution of the nanocrystal hosts, the position distribution of erbium spins (**Figure S12**), and the Poisson distribution in the number of erbium spins per nanocrystal host. Further, the effects of instantaneous diffusion is sensitive to the fraction of flipped spins[56] and can be determined by the $\pi$-pulse and the inhomogeneous distribution of erbium spins measured in field-swept echo detected spectra (**Methods**, **Figure S14**). Altogether, we observe reasonable agreement between experimental data and theoretical simulations of $T_2$, suggesting that CCE calculations are an appropriate framework to analyze coherence times even in nanocrystal hosts. The differences in calculated coherence times at high and low concentrations can likely be attributed to the following: at high concentrations there is a stronger effect of flipping an even smaller fraction of spins with the $\pi$-pulse, due to rapid dephasing during Rabi oscillations and broader inhomogeneous distribution. At low concentrations, where the spin coherence should be dominated by the hydrogen nuclear spin bath, we already observed that slight differences in the erbium defect position distribution can yield very different coherence times (**Figure 5a**). As a result, we hypothesize that erbium spins may be preferentially closer to the surface, more so than the amount given by the natural density of atoms in a nanocrystal (**Figure S12**). This viewpoint is consistent with the exclusion of dopants within the critical nuclei of a nanocrystal[58]. The other possibility for the coherence time discrepancy may be due to paramagnetic species that were not accounted for in these simulations, which likely consists of either $Ce^{3+}$ or $O^{2-}$ complexes, as observed earlier in STEM-EELS (**Figure 2d**). $Ce^{3+}$ is likely at least partially coordinated with the oleate ligands as part of the nanocrystal synthesis but may still act as a source of magnetic noise for erbium spins, depending on its own spin relaxation dynamics and coordination environment. Understanding the specific relaxation dynamics as well as the concentration of paramagnetic spins in the surrounding bath will be important to further understand the limits to spin coherence[8]. For example, we find that $Er^{3+}$ spin coherence can be limited by indirect flip-flops[59,60] or by the $T_1$-induced spin flips[61] of nearby paramagnetic spins, depending on the concentration of extraneous paramagnetic spins (**Figure S15**). Based on these calculations we can safely constrain the number of extraneous surface spins to less than ten per nanoparticle. With further experimental studies, it should be possible to pinpoint the exact effects of $Ce^{3+}$ impurities on $Er^{3+}$ decoherence.

More generally, we expect that intraparticle interactions between $Er^{3+}$ within a nanocrystal to dominate the spin dynamics (i.e., relaxation, coherence, linewidth) whenever there is more than approximately a single $Er^{3+}$ per nanocrystal. Specifically, intraparticle $Er^{3+}$ interactions have both large gyromagnetic ratios ($\gamma_{Er3+} = 95$ MHz/mT, compared to $\gamma_{1H} = 0.043$ MHz/mT and $\gamma_{e-} = 28$ MHz/mT) and the dipolar interactions are resonant, so that $Er^{3+}$ flip-flop interactions cannot be neglected. Therefore, effects such as the size-dependent coherence time calculated in Figure 5b will likely not be observable in most systems except when $\langle N_{Er} \rangle \sim 1$. A general method to improve the spin coherence in these systems is to estimate the dephasing rates of possible limiting mechanisms and reduce those rates until a new limiting mechanism is revealed. This scheme can be viewed analogously to the work done in improving the photoluminescence quantum yield of semiconductor systems[14,62], where different nonradiative mechanisms are revealed as the material quality is improved. In this work, the concentration of $Er^{3+}$ was reduced until the spin coherence was no longer limited by instantaneous diffusion, but rather, by surface effects. Clear modulations of the spin echo signal by hydrogen nuclei demonstrates that it is a large contributing factor to the limiting mechanism. Yet, it is perhaps surprising that paramagnetic surface noise does not limit the spin coherence from the outset, considering the proximity of $Er^{3+}$ to the surface and the prevalence of $Ce^{3+}$ and oxygen vacancies in ceria systems[63,64]. Understanding the structural, chemical, and spin properties of the nanocrystal surface will therefore be imperative for further development in these systems, analogous to the previous work done in colloidal quantum dot science[65,66] and shallow NV centers in diamond[13,67].

**Conclusions**

Our results demonstrate that optically active erbium spin defects in cerium dioxide nanocrystal hosts represent an exciting platform for developing quantum hardware. The quantum coherence observed here, resulting from a unique combination of $Er^{3+}$ ions in cubic symmetry embedded in a nuclear-spin free $CeO_2$ nanocrystal host, should enable opportunities in both quantum sensing and quantum communications. For example, despite the relatively short spin coherence times of $Er^{3+}$ ions, their proximity to the surface combined with their large gyromagnetic ratios should enable magnetic field sensitivities comparable to or even better than shallow NV centers in diamond[12]. Utilizing core-shell structures to improve spin coherence times and modifying the nanocrystal surfaces should facilitate sensitivity down to a single molecule. Dynamical decoupling pulse sequence engineering can also be utilized to reduce sensitivity to specific types of noises, such as those dipolar in character or low frequency[68,69]. Finally, given the unique optical properties of $Er^{3+}$, there may be opportunities to embed these nanocrystals into a photonic structure[11,70], significantly improving their spin-photon interface. Quantum memory applications may therefore be possible with sufficient improvement in the $Er^{3+}$ ions' optical and spin coherence.

More broadly, our work demonstrates the unique challenges and tools needed to fully understand spin defects in nanocrystal hosts, as their surfaces are deeply entangled with their properties. For example, deterministically placing and measuring the atomic position of quantum defects in nanocrystal hosts will likely be critical to their coherence properties. Moreover, a complete understanding of extraneous surface noise will require a combination of extensive surface spectroscopy and single defect measurements. Yet, there should be many opportunities to both significantly improve their quantum coherence and develop new quantum systems with distinct functionality, especially considering the thermodynamic and kinetic factors that favor few extraneous defects and synthetic tunability in colloidal nanocrystals. For example, the substantially reduced instantaneous diffusion given by the finite size of the nanocrystal could enable opportunities for studying quantum coherent many-body systems. Furthermore, synthesizing and controlling systems at the limit of a single defect should in principle be simpler and more modular in nanocrystal hosts, as long as the surface-associated decoherence can be limited. Unique to this science will therefore be a strong understanding of the surface and how to mitigate its impact on quantum coherence, which will likely be broadly useful for many systems beyond the one investigated here.

**Methods**

**Materials**

Water (Sigma-Aldrich, ACS reagent, for ultratrace analysis), hydrogen peroxide (Sigma Aldrich, 30% (W/W), for ultratrace analysis), nitric acid (Sigma-Aldrich, 70% redistilled, >99.999%), oleic acid (Fisher-Scientific, >99%), tert-butylamine (Sigma-Aldrich, >99.5%), cerium (III) nitrate hexahydrate (Sigma-Aldrich, 99.999%), erbium (III) nitrate pentahydrate (Sigma-Aldrich, 99.9%), toluene (Sigma-Aldrich, 99.8%, anhydrous), hexane (Sigma-Aldrich, 95%, anhydrous), methanol (Sigma-Aldrich, 99.8%, anhydrous) were used without subsequent purification.

**Synthesis of Erbium Doped Ceria Nanocrystals**

We followed a synthetic method similar to that for pure ceria nanocrystal synthesis[71] with slight modifications to yield erbium doped nanocrystals. Briefly, we prepared a 16.7 mmol/L stock solution of aqueous cerium nitrate (~145 mg in 20 mL) and erbium nitrate (~148 mg in 20 mL). Then, a specific quantity of the cerium and erbium nitrate stock solutions were pipetted depending on the desired molar ratio of erbium to cerium. For example, a 20000 PPM (2%) erbium sample would require 7353 µL of the aqueous cerium nitrate solution and 147 µL of the aqueous erbium nitrate solution. For lower concentrations, it was useful to perform serial dilutions of the erbium stock solution to reduce errors in pipetting small volumes. The two aqueous solutions were mixed and subsequently added to a 20 mL sized Teflon-lined autoclave, to which 7.5 mL of toluene, 750 µL of oleic acid, and 75 µL of tert-butylamine were added without stirring. The autoclave was heated in an oven set to 180 ºC for 24 hours, which was allowed to cool naturally to room temperature. The nanocrystals were then flocculated with methanol (~2 - 3 mL), centrifuged at 10000 RPM for 5 minutes, and finally dispersed in nonpolar solvents such as toluene or hexane. Repeated flocculation and redispersion was performed as necessary to remove unreacted precursors, contaminants, and side products, which could be monitored via ICP-OES.

**Electron Paramagnetic Resonance Spectroscopy**

Electron paramagnetic resonance (EPR) measurements were performed by filling a 4 mm OD Wilmad quartz tube (Sigma-Aldrich) in a nitrogen atmosphere with erbium doped ceria nanocrystals dissolved in toluene solution with a concentration of either 5 mg/mL for the $\langle N_{Er} \rangle$ = 0.3, 1.2 samples, 50 µg/mL for the $\langle N_{Er} \rangle$ = 3.2 sample, and 10 µg/mL for all other samples. CW Measurements were performed at X-band frequency (~9.5 GHz) in a Bruker ELEXSYS II E500 spectrometer (Bruker BioSpin, Ettlingen, Germany) employing a $TE_{102}$ rectangular EPR resonator (Bruker ER4102ST). Measurements were performed at approximately 3.2K unless otherwise noted using a flow cryostat (ICE Oxford, UK) with pumped liquid helium. At this temperature, toluene forms a glassy matrix embedded with nanocrystals. Temperature dependent measurements were performed using an ITC temperature controller (Oxford Instruments, UK). Measurements were performed at sufficiently high microwave powers to ensure good signal to noise ratios while avoiding saturation-induced linewidth broadening.

Pulsed EPR measurements were also performed at X-band frequency (~9.7 GHz) in a Bruker ELEXSYS E580 spectrometer (Bruker BioSpin, Ettlingen, Germany) that is equipped with a dielectric ring resonator (Bruker ER 4118-MD5-W1). An CF935 cryostat in combination with an ITC temperature controller (both Oxford Instruments, UK) was used. Electron spin echoes were first detected and a $\pi$-pulse length of 24 ns was fixed. The microwave power was then adjusted to maximize the spin echo intensity. The corresponding $\frac{\pi}{2}$ pulse is 12 ns. Field-swept electron spin-echo detected EPR spectra were recorded using a two-pulse Hahn-echo sequence with a 12 ns $\pi$/2-pulse, followed by a 24 ns $\pi$-pulse with a fixed interpulse delay $\tau_0$ of either 200 ns for the $\langle N_{Er} \rangle = 0.3$ sample, or 110 ns for all other samples. For the spin coherence measurements, the same two-pulse sequence was used except with a varying delay time $\tau$. The typical measurement deadtime was ~100 ns. For inversion recovery spin relaxation measurements, a three-pulse inversion sequence was used where a 24 ns $\pi$-pulse (inversion pulse) is followed by a two-pulse Hahn echo sequence with varying time delay $T$ between the inversion $\pi$-pulse and the Hahn echo sequence. The echo detected sequence used here is the same as that in the field-swept electron spin-echo detected spectra, i.e., a microwave pulse sequence composed of a 12 ns $\pi/2$-pulse followed by a 24 ns $\pi$-pulse with a fixed delay $\tau_0$ of either 200 ns for the $\langle N_{Er} \rangle = 0.3$ sample, or 110 ns for all other samples. Microwave power was adjusted to maximize the echo intensity and the Davies integral criterion was used to maximize the signal-to-noise ratio for echo detection.

**Transmission Electron Microscopy Measurements**

Nanocrystal morphology and size distributions were characterized by first diluting the native solutions to roughly 50 μg/mL concentration, to which roughly 1 μL of solution was dropped onto a 3 mm TEM grid with either a ~5-6 nm carbon film (CF400-Cu) for morphology and size distribution measurements or lacey carbon film (<3 nm thick) (LC400-Au-CC) for high resolution TEM or STEM imaging. Samples were dried before measuring with a 200-kV FEI Tecnai G2 T20 S-TWIN microscope with a Gatan RIO camera at 4096 by 4096 resolution with magnification factors up to 450 kX. Images were acquired with typical exposure times from 1 to 4 seconds at a variety of magnification ratios to ascertain their size and shape distributions.

**Scanning Transmission Electron Microscopy and Electron Energy Loss Spectroscopy Measurements**

Atomic-resolution characterization of the erbium doped ceria nanocrystals was conducted using the aberration-corrected scanning transmission electron microscope (STEM) JEOL ARM200CF at the University of Illinois Chicago, operated at 200 kV. The microscope is equipped with a cold field emission gun and a CEOS aberration corrector resulting in a probe size of 0.78 Å with 350 meV energy resolution. A probe convergence angle of 30 mrad was used for STEM imaging experiments. The inner and outer collection angles for high-angle annular dark-field (HAADF) imaging were set to 90 mrad and 370 mrad respectively. The inner and outer collection angles for annular bright-field (ABF) imaging were set to 11 mrad and 23 mrad respectively. To reduce electron-beam induced damage, the dose threshold was kept below 800 e/Å$^2$. Atomic-resolution images were acquired sequentially, where consecutive frames (10-20) were aligned and integrated to obtain high-quality images. Electron energy loss (EELS) experiments were performed using a dual-range Gatan Continuum spectrometer with an entrance aperture of 5 mm and an energy dispersion of 0.3 eV/channel. The O K and Ce M edges were acquired with an acquisition time of 0.5 seconds, while Er M edge was acquired simultaneously with an acquisition time of 2 seconds. We have performed principal component analysis to improve the signal-to-noise ratio of the EEL spectra. A power law background model was used prior to the O K, Ce M and Er M edges. Energy dispersive X-ray spectroscopy (EDS) was performed using an Oxford X-Max 100LTE windowless silicon drift X-ray detector. Ceria nanocrystals were prepared for STEM analysis by first diluting to roughly 50 μg/mL concentration, and then drop-casting on a TEM grid. Samples were heated at 80 °C for 2 hours prior to STEM experiments to reduce surface contamination.

**Powder X-ray Diffraction Measurements**

Nanocrystal solutions were drop-cast onto a single crystal silicon wafer with low background (Bruker AXS) from high vapor point solvents such as hexane. They were allowed to dry quickly to prevent superlattice formation of the nanocrystal powder and instead produce randomly oriented nanocrystals. Measurements were done on a Bruker D2 Phaser instrument and samples were irradiated with copper K-alpha X-rays with a wavelength of 1.5418 A. Spectra were collected at angles between 10° and 80° $2\theta$.

**Inductively Coupled Plasma Optical Emission Spectroscopy and Mass Spectroscopy**

Samples prepared for ICP-OES or ICP-MS followed a protocol similar to that developed by Buhro et al[72]. Briefly, a nanocrystal solution of 50 µL was micro-pipetted into a centrifuge tube. To the samples, 500 µL of 30% $H_2O_2$ was added to the samples and the centrifuge tube was immediately capped. After a few minutes, 500 µL of 70% $HNO_3$ was added, and the centrifuge tube was also immediately capped. Samples were left to digest overnight in the hydrogen peroxide/nitric acid solution. The next morning, 9 mL of DI $H_2O$ was added to the solution to be used for ICP-OES. An ideal target concentration for erbium was either 5 to 500 ng/mL for ICP-MS, or 500 ng/mL to 100 µg/mL for ICP-OES. Digested sample solutions were serial diluted to further reduce their concentrations if necessary, and multiple solution concentrations were generated for the same sample to confirm the erbium to cerium ratio. Cerium and Erbium standards were purchased from Agilent (1000 µg/mL) and a standard series of 100 µg/mL, 25 µg/mL, 5 µg/mL, 1 µg/mL, 200 ng/mL, 40 ng/mL, and 8 ng/mL were typically made by the addition of $H_2O_2$, $HNO_3$, DI $H_2O$, and an appropriate amount of serial dilution. ICP-OES and ICP-MS measurements were done in an Agilent 700 Series spectrometer and Agilent 7700x Series, respectively. A diluted solution of 2% HCl and 1% $HNO_3$ was used to rinse the system between samples.

**Photoluminescence Excitation Spectroscopy Measurements**

The optical data is measured using a confocal microscopy setup. The sample is mounted in a close-cycle cryostat (Montana Instrument, s50) with a base temperature ~3.3 K and a sample temperature ~ 3.6 K. A continuously tunable C-band laser with a 1460 nm-1570 nm tuning range (Toptica CTL 1500) is used for optical excitation. The laser is coupled to a single-mode fiber and sent through multiple optical modulators to create optical excitation pulses of various length and frequency modulation depending on the requirements of various types of optical measurements. For photoluminescence excitation (PLE) measurements, the laser light is modulated by two fiber-based acousto-optic modulators (AOM, AA opto-electronic) with a rise time of about 30 ns and an extinction >45 dB to generate clean 1.5 ms excitation pulses. The modulated laser is sent through a 50/50 beamsplitter and then focused on the sample using an infrared objective with NA 0.65. Photoluminescence (PL) from the sample is collected by the same objective, passed through the 50/50 beamsplitter, filtered by two 1500 nm long pass filters, and then coupled into a single mode fiber connected to superconducting nanowire single-photon detector (SNSPD, Quantum Opus) for detection. A fiber-based acousto-optic modulator (AOM, AA opto-electronic) with a rise time of about 30 ns is attached to a single-mode fiber to time-gate the photon detection. The AOM on the collection side, as well as the AOM on the excitation side is controlled by the data acquisition system (DAQ, National Instruments) to set up the properly aligned sequence of collection window with respect to excitation sequences. A collection window of 7 ms opening right after the end of the 1.5 ms optical excitation pulse is used to time-gate and collect the PL from $Er^{3+}$ ions for PLE measurements.

**Analysis**

**Nanocrystal size distribution**

Size distributions were calculated using either home-written codes or the recently developed autoDetectMNP code[73] to ascertain an estimated projected area $A_{proj}$ per nanocrystal. Projected areas were then converted into projected spherical diameters $d_{proj}$ by using $d_{proj} = \sqrt{\frac{4}{\pi} A_{proj}}$ and the histogrammed distribution was then used to calculate an average diameter $d$ for each nanocrystal.

**Estimation of the number of erbium defects per nanocrystal**

Ensemble average erbium dopants per nanocrystal were estimated by using the following analysis. First, the ceria nanocrystal size distribution was ascertained from TEM measurements, which gave an estimated average diameter $d$ for each nanocrystal. The total volume of the nanocrystal in a spherical projection would then be $\frac{\pi}{6}d^3$. The total number of cerium atoms per nanocrystal was then estimated by assuming a unit cell lattice constant of 5.42 Angstroms for which 4 cerium atoms would be embedded in, so that the average volume per cerium atom would be 39.8 cubic Angstroms. The total cerium atoms per nanocrystal was then calculated by dividing the total volume of the nanocrystal ($\frac{\pi}{6}d^3$) by the average volume per cerium atom (39.8 cubic Angstroms). The average erbium concentration for the ensemble would then be the total cerium atoms per nanocrystal, multiplied by the erbium to cerium ratio ascertained via either ICP-OES or ICP-MS, i.e.,

$$\langle N_{Er} \rangle = \frac{\pi}{6}d_{proj}^3 \frac{4}{a^3}\rho_{Er}$$

Where $\rho_{Er}$ describes the molar ratio of erbium to cerium. The largest error bars in this estimation pertain to the finite size distribution of the nanocrystals, given its cubic dependence.

**Electron paramagnetic resonance spectra and dynamics**

EPR spectra were first normalized, before using EasySpin[74] (6.0.0) in MATLAB (R2023b) to perform matrix calculations and fitting. A linear background was used. Briefly, the Hamiltonian considered for this spin system is given as

$$\hat{H} = \hat{H}_{EZI} + \hat{H}_{NZI} + \hat{H}_{HFI}$$

where

$$\hat{H}_{EZI} = \mu_B \boldsymbol{B} \boldsymbol{g} \boldsymbol{S}$$

is the electron Zeeman interaction in a magnetic field $\boldsymbol{B}$, with an electron g-tensor $\boldsymbol{g}$ and an electron spin $\boldsymbol{S}$. Here, $\mu_B$ is the electron Bohr magneton. Similarly,

$$\hat{H}_{NZI} = -\mu_n g_n \boldsymbol{B} \boldsymbol{I}$$

is the nuclear Zeeman interaction in a magnetic field $\boldsymbol{B}$ interacting with a nuclear spin $\boldsymbol{I}$, with nuclear g-factor $g_n$ and nuclear magneton $\mu_n$. Further, we have neglected the nuclear quadrupolar interaction, which is expected to vanish in a cubic symmetric site[75]. We consider a spin S=$\frac{1}{2}$ system for $Er^{3+}$ and an isotropic $g$ factor ($g_{xx} = g_{yy} = g_{zz}$) given the cubic symmetry of $Er^{3+}$. The hyperfine interaction is given as

$$\hat{H}_{HFI} = \boldsymbol{S} \boldsymbol{A} \boldsymbol{I}$$

Where $\boldsymbol{S}$ is the electron spin vector with spin ½, $\boldsymbol{A}$ is the hyperfine interaction tensor, and $\boldsymbol{I}$ is the nuclear spin vector with a nuclear spin of $\frac{7}{2}$ from the interaction with the $^{167}$Er nuclei, which is 22.9% naturally abundant. We consider a hyperfine interaction tensor that is isotropic given the cubic site symmetry. We further considered separate linewidths for the $I = 0$ and $I = \frac{7}{2}$ nuclear spin populations, and combined their broadened spectra weighted by the natural abundance of $^{167}$Er. Fits were performed using a Voigt function and total linewidths were calculated using the approximate expression[76]

$$f_V = 0.5346 f_L + \sqrt{0.2166 f_L^2 + f_G^2}$$

from the individual Gaussian and Lorentzian contributions. The electron g factor and nuclear hyperfine interaction were allowed to vary to achieve the best possible fit. The power law fitting of the dipolar linewidths and the temperature dependent linewidth fits were performed in MATLAB using a least-squares curve fit.

For the spin-echo dynamics (Hahn echo decays), we fitted to either a typical exponential decay, i.e., $\sim \exp(-2\tau/T_2)$, or to the Fourier transform of a shifted Lorentzian model, i.e.,

$$S(f) = \sum_i a_i \frac{\frac{\Gamma}{2}}{(f - f_i)^2 + \left(\frac{\Gamma}{2}\right)^2}$$

where $\Gamma = \frac{1}{\pi T_2}$ and the $2\tau$ in the exponent must be properly accounted for when examining $T_2$ and the corresponding hyperfine couplings. This model is equivalent to the typical exponential decay in the limit where $f_i \rightarrow 0$. In the simplest case with a single external frequency ($^1$H Larmor frequency), the above expression reduces to a decaying sinusoidal exponential.

For the inversion recovery dynamics, we fitted the dynamics to a multiexponential decay $\sum_i A_i \exp\left(-\frac{T}{T_i}\right)$. Given the large range in rate constants for the low erbium dopant nanocrystals, inversion recovery dynamics were additionally measured with short time delays and stitched together with the long-time delays, as detailed in **Figure S10**.

**Theoretical calculations of the spin coherence**

Cluster Correlation Expansion (CCE) simulations were carried out with the PyCCE package[77]. In the CCE framework, the coherence function $\mathcal{L}(t)$ is equal to the normalized element of the central spin density matrix, and is factorized into the product of contributions of clusters of different size:

$$\mathcal{L}(t) \approx \prod_{C \subseteq \{1,2,...n\}} \tilde{L}_C(\text{t})$$

Where $\tilde{L}_C$ are irreducible contribution of the cluster $C$, and $n$ is the total number of clusters included in the expansion. The maximum size of the cluster determines the order of the approximation.

We consider a system consisting of a central electron spin-½ and hydrogen spins ½ in an external magnetic field; the Hamiltonian is given by:

$$\hat{H} = -\mu_B g_{zz} B_z \hat{S}_z - \sum_i \gamma_n B_z \hat{I}_{iz} + \sum_i \boldsymbol{S} \boldsymbol{A} \, \boldsymbol{I}_i + \sum_{i \neq j} \boldsymbol{I}_i \boldsymbol{P} \boldsymbol{I}_j$$

where $\boldsymbol{S}$ is the central spin, $\boldsymbol{I}_i$ are the hydrogen spins. The $\boldsymbol{A}$ tensor denotes the hyperfine between the central spin and the hydrogens. The $\boldsymbol{P}$ tensor is the spin dipole-dipole intrabath interaction between the bath spins.

To model the hydrogen nuclear spin bath, we construct an idealized spherical nanoparticle with the hydrogen nuclear spins randomly distributed around it. We take the concentration of the hydrogen to be the same as in oleic acid; oleic acid almost completely covers the surface of nanoparticles, and the difference in hydrogen concentration in toluene and oleic acid should not impact the qualitative results of our work.

Calculations were carried out at third order for erbium in the center of the nanoparticle, and at second order at varied distance from the surface.

To model the instantaneous diffusion, we numerically generate 100000 random distributions of erbium ions in the nanoparticles at a given concentration, and compute the dipolar broadening, induced by the other erbiums atoms on a single ion as:

$$\gamma = \frac{1}{2\sqrt{2}}\sqrt{\sum_i U_{zz}^2}$$

The distribution of the broadening obtained in our calculations is shown in **Figure 5c** of the main text. Next, we numerically integrate over this distribution. The coherence time $T_2$ is the inverse of the mean decoherence rate multiplied by the fraction $f$ of the erbium spins, flipped by the $\pi$-pulse to obtain the final decoherence rate $T_2 = (f\bar{\gamma})^{-1}$. The fraction $f$ is obtained by integrating over the experimental erbium linewidth $\rho(\delta)$, given by the experimental Rabi frequency $\omega_R$ and pulse length $\tau$:

$$f = \int_{-\infty}^{+\infty} \frac{\omega_R^2}{\delta^2 + \omega_R^2} \sin^2\left(\frac{\tau}{2}\sqrt{\delta^2 + \omega_R^2}\right)\rho(\delta)d\delta$$

To estimate the adverse effect of $Ce^{3+}$ paramagnetic impurities on the $Er^{3+}$ electron spin coherence, we simulate a model system of $Ce^{3+}$ as electron spins with $S = 1/2$ and $g = 2$, randomly placed on the surface of a 7 nm diameter nanosphere. $Er^{3+}$ is placed in the center of the nanosphere. We vary the number of electron spins from 1 to 100 and vary their intrinsic relaxation rate, $T_1$, between 10 ns and 10 ms. We use Master equation CCE (ME-CCE)[78] in the second order to account for the finite relaxation rate of $Ce^{3+}$, stemming from interactions with unknown Markovian noise sources (e.g., phononic bath, charge fluctuations, and others).

### Acknowledgements

The authors would like to thank T.X. Zheng, P. Maurer, N. Delegan, W. Baek for useful discussions and constructive feedback, H.C. Fry for assistance with EPR measurements at CNM, J. Li for assistance with ICP-MS measurements, X. Wang for assistance with TEM analysis, F. Shi for assistance with STEM measurements, and J. Jureller for assistance with MRSEC facilities and instrumentation. We acknowledge the MRSEC Shared User Facilities at the University of Chicago (NSF DMR-2011854). Work performed at the Center for Nanoscale Materials, a U.S. Department of Energy Office of Science User Facility, was supported by the U.S. DOE, Office of Basic Energy Sciences, under Contract No. DE-AC02-06CH11357. The computational study (M. Onizhuk, J. Nagura, G. Galli) was supported by AFOSRFA9550-22-1-0370. We acknowledge the use of the computational facilities (Research Computing Center) at the University of Chicago. The development of the PyCCE code used in this work was supported by the Computational Materials Science Center Midwest Integrated Center for Computational Materials (MICCoM). MICCoM is part of the Computational Materials Sciences Program funded by the U.S. Department of Energy, Office of Science, Basic Energy Sciences, Materials Sciences, and Engineering Division through the Argonne National Laboratory, under Contract No. DE-AC02-06CH11357. A.S. Thind and R.F. Klie acknowledge funding support from the National Science Foundation grant NSF DMR2309396. The EPR work in the Chemical Sciences and Engineering Division at Argonne National Laboratory (J.K. Bindra, J. Niklas, O.G. Poluektov) was supported by the U.S. Department of Energy, Office of Science, Office of Basic Energy Sciences, Division of Chemical Sciences, Geosciences, and Biosciences, through Argonne National

Laboratory under Contract No. DE-AC02-06CH11357. Work at Argonne was primarily funded (C. Wicker, J. Zhang, F. J. Heremans, D. D. Awschalom) by the U.S. Department of Energy, Office of Science, Basic Energy Sciences, Materials Sciences and Engineering Division, including support for optical and spin characterization studies. We acknowledge additional support (G.D. Grant) from Q-NEXT, a U.S. Department of Energy Office of Science National Quantum Information Science Research Centers under Award Number DE-FOA-0002253.

**Author Contributions**

J. Wong conceived the main ideas for the project. J. Wong performed synthesis of the nanocrystals and led the characterization and data analysis effort on the nanocrystals, including their structural, chemical, optical, and spin properties. M. Onizhuk and J. Nagura performed coherence calculations, supervised by G. Galli. A. S. Thind performed aberration-corrected STEM-EELS measurements, supervised by R.F. Klie. J.K. Bindra, J. Zhang, J. Niklas, and O.G. Poluektov assisted with CW and pulsed EPR measurements. C. Wicker, J. Zhang, and G.D. Grant aided in optical measurements and interpretation of results, supervised by F.J. Heremans, and D.D. Awschalom. Y. Zhang aided with nanocrystal synthesis and spin characterization. A.P. Alivisatos supervised the project and provided scientific input at all stages. J. Wong wrote the manuscript with input from all authors.

**ORCID**

Joeson Wong: 0000-0002-6304-7602

Mykyta Onizhuk: 0000-0003-0434-4575

Jonah Nagura: 0000-0002-2377-9702

Arashdeep Singh Thind: 0000-0003-0262-766X

Jasleen K. Bindra: 0000-0001-7031-7482

Yuxuan Zhang: 0000-0001-8664-017X

Jens Niklas: 0000-0002-6462-2680

Oleg G. Poluektov: 0000-0003-3067-9272

Robert F. Klie: 0000-0003-4773-6667

Jiefei Zhang: 0000-0002-7329-3110

Giulia Galli: 0000-0002-8001-5290

F. Joseph Heremans: 0000-0003-3337-7958

David D. Awschalom: 0000-0002-8591-2687

A. Paul Alivisatos: 0000-0001-6895-9048

**Competing Interests**

The authors declare no competing financial interest.

## References

(1) Degen, C. L.; Reinhard, F.; Cappellaro, P. Quantum Sensing. *Rev. Mod. Phys.* **2017**, *89* (3), 035002. https://doi.org/10.1103/RevModPhys.89.035002.

(2) Giovannetti, V.; Lloyd, S.; Maccone, L. Advances in Quantum Metrology. *Nature Photon* **2011**, *5* (4), 222–229. https://doi.org/10.1038/nphoton.2011.35.

(3) Kimble, H. J. The Quantum Internet. *Nature* **2008**, *453* (7198), 1023–1030. https://doi.org/10.1038/nature07127.

(4) Gisin, N.; Thew, R. Quantum Communication. *Nature Photon* **2007**, *1* (3), 165–171. https://doi.org/10.1038/nphoton.2007.22.

(5) Preskill, J. Quantum Computing in the NISQ Era and Beyond. *Quantum* **2018**, *2*, 79. https://doi.org/10.22331/q-2018-08-06-79.

(6) Georgescu, I. M.; Ashhab, S.; Nori, F. Quantum Simulation. *Rev. Mod. Phys.* **2014**, *86* (1), 153–185. https://doi.org/10.1103/RevModPhys.86.153.

(7) de Leon, N. P.; Itoh, K. M.; Kim, D.; Mehta, K. K.; Northup, T. E.; Paik, H.; Palmer, B. S.; Samarth, N.; Sangtawesin, S.; Steuerman, D. W. Materials Challenges and Opportunities for Quantum Computing Hardware. *Science* **2021**, *372* (6539), eabb2823. https://doi.org/10.1126/science.abb2823.

(8) Wolfowicz, G.; Heremans, F. J.; Anderson, C. P.; Kanai, S.; Seo, H.; Gali, A.; Galli, G.; Awschalom, D. D. Quantum Guidelines for Solid-State Spin Defects. *Nat Rev Mater* **2021**, *6* (10), 906–925. https://doi.org/10.1038/s41578-021-00306-y.

(9) Awschalom, D. D.; Hanson, R.; Wrachtrup, J.; Zhou, B. B. Quantum Technologies with Optically Interfaced Solid-State Spins. *Nature Photon* **2018**, *12* (9), 516–527. https://doi.org/10.1038/s41566-018-0232-2.

(10) Guo, X.; Delegan, N.; Karsch, J. C.; Li, Z.; Liu, T.; Shreiner, R.; Butcher, A.; Awschalom, D. D.; Heremans, F. J.; High, A. A. Tunable and Transferable Diamond Membranes for Integrated Quantum Technologies. *Nano Lett.* **2021**, *21* (24), 10392–10399. https://doi.org/10.1021/acs.nanolett.1c03703.

(11) Raha, M.; Chen, S.; Phenicie, C. M.; Ourari, S.; Dibos, A. M.; Thompson, J. D. Optical Quantum Nondemolition Measurement of a Single Rare Earth Ion Qubit. *Nat Commun* **2020**, *11* (1), 1605. https://doi.org/10.1038/s41467-020-15138-7.

(12) Yu, C.-J.; von Kugelgen, S.; Laorenza, D. W.; Freedman, D. E. A Molecular Approach to Quantum Sensing. *ACS Cent. Sci.* **2021**, *7* (5), 712–723. https://doi.org/10.1021/acscentsci.0c00737.

(13) Sangtawesin, S.; Dwyer, B. L.; Srinivasan, S.; Allred, J. J.; Rodgers, L. V. H.; De Greve, K.; Stacey, A.; Dontschuk, N.; O'Donnell, K. M.; Hu, D.; Evans, D. A.; Jaye, C.; Fischer, D. A.; Markham, M. L.; Twitchen, D. J.; Park, H.; Lukin, M. D.; de Leon, N. P. Origins of Diamond Surface Noise Probed by Correlating Single-Spin Measurements with Surface Spectroscopy. *Phys. Rev. X* **2019**, *9* (3), 031052. https://doi.org/10.1103/PhysRevX.9.031052.

(14) Hanifi, D. A.; Bronstein, N. D.; Koscher, B. A.; Nett, Z.; Swabeck, J. K.; Takano, K.; Schwartzberg, A. M.; Maserati, L.; Vandewal, K.; van de Burgt, Y.; Salleo, A.; Alivisatos, A. P. Redefining Near-Unity Luminescence in Quantum Dots with Photothermal Threshold Quantum Yield. *Science* **2019**, *363* (6432), 1199–1202. https://doi.org/10.1126/science.aat3803.

(15) Utzat, H.; Sun, W.; Kaplan, A. E. K.; Krieg, F.; Ginterseder, M.; Spokoyny, B.; Klein, N. D.; Shulenberger, K. E.; Perkinson, C. F.; Kovalenko, M. V.; Bawendi, M. G. Coherent Single-Photon Emission from Colloidal Lead Halide Perovskite Quantum Dots. *Science* **2019**, *363* (6431), 1068–1072. https://doi.org/10.1126/science.aau7392.

(16) Koscher, B. A.; Swabeck, J. K.; Bronstein, N. D.; Alivisatos, A. P. Essentially Trap-Free CsPbBr3 Colloidal Nanocrystals by Postsynthetic Thiocyanate Surface Treatment. *J. Am. Chem. Soc.* **2017**, *139* (19), 6566–6569. https://doi.org/10.1021/jacs.7b02817.

(17) Won, Y.-H.; Cho, O.; Kim, T.; Chung, D.-Y.; Kim, T.; Chung, H.; Jang, H.; Lee, J.; Kim, D.; Jang, E. Highly Efficient and Stable InP/ZnSe/ZnS Quantum Dot Light-Emitting Diodes. *Nature* **2019**, *575* (7784), 634–638. https://doi.org/10.1038/s41586-019-1771-5.

(18) McDonald, S. A.; Konstantatos, G.; Zhang, S.; Cyr, P. W.; Klem, E. J. D.; Levina, L.; Sargent, E. H. Solution-Processed PbS Quantum Dot Infrared Photodetectors and Photovoltaics. *Nature Mater* **2005**, *4* (2), 138–142. https://doi.org/10.1038/nmat1299.

(19) Kamat, P. V. Quantum Dot Solar Cells. Semiconductor Nanocrystals as Light Harvesters. *J. Phys. Chem. C* **2008**, *112* (48), 18737–18753. https://doi.org/10.1021/jp806791s.

(20) Kim, J.; Kim, H. S.; Lee, N.; Kim, T.; Kim, H.; Yu, T.; Song, I. C.; Moon, W. K.; Hyeon, T. Multifunctional Uniform Nanoparticles Composed of a Magnetite Nanocrystal Core and a Mesoporous Silica Shell for Magnetic Resonance and Fluorescence Imaging and for Drug Delivery. *Angewandte Chemie International Edition* **2008**, *47* (44), 8438–8441. https://doi.org/10.1002/anie.200802469.

(21) Alivisatos, P. The Use of Nanocrystals in Biological Detection. *Nat Biotechnol* **2004**, *22* (1), 47–52. https://doi.org/10.1038/nbt927.

(22) Parak, W. J.; Gerion, D.; Pellegrino, T.; Zanchet, D.; Micheel, C.; Williams, S. C.; Boudreau, R.; Gros, M. A. L.; Larabell, C. A.; Alivisatos, A. P. Biological Applications of Colloidal Nanocrystals. *Nanotechnology* **2003**, *14* (7), R15. https://doi.org/10.1088/0957-4484/14/7/201.

(23) Hoang, T. B.; Akselrod, G. M.; Mikkelsen, M. H. Ultrafast Room-Temperature Single Photon Emission from Quantum Dots Coupled to Plasmonic Nanocavities. *Nano Lett.* **2016**, *16* (1), 270–275. https://doi.org/10.1021/acs.nanolett.5b03724.

(24) Chen, Y.; Ryou, A.; Friedfeld, M. R.; Fryett, T.; Whitehead, J.; Cossairt, B. M.; Majumdar, A. Deterministic Positioning of Colloidal Quantum Dots on Silicon Nitride Nanobeam Cavities. *Nano Lett.* **2018**, *18* (10), 6404–6410. https://doi.org/10.1021/acs.nanolett.8b02764.

(25) Xie, C.; Niu, Z.; Kim, D.; Li, M.; Yang, P. Surface and Interface Control in Nanoparticle Catalysis. *Chem. Rev.* **2020**, *120* (2), 1184–1249. https://doi.org/10.1021/acs.chemrev.9b00220.

(26) Almutlaq, J.; Liu, Y.; Mir, W. J.; Sabatini, R. P.; Englund, D.; Bakr, O. M.; Sargent, E. H. Engineering Colloidal Semiconductor Nanocrystals for Quantum Information Processing. *Nat. Nanotechnol.* **2024**, 1–10. https://doi.org/10.1038/s41565-024-01606-4.

(27) Kagan, C. R.; Bassett, L. C.; Murray, C. B.; Thompson, S. M. Colloidal Quantum Dots as Platforms for Quantum Information Science. *Chem. Rev.* **2021**, *121* (5), 3186–3233. https://doi.org/10.1021/acs.chemrev.0c00831.

(28) Fainblat, R.; Barrows, C. J.; Gamelin, D. R. Single Magnetic Impurities in Colloidal Quantum Dots and Magic-Size Clusters. *Chem. Mater.* **2017**, *29* (19), 8023–8036. https://doi.org/10.1021/acs.chemmater.7b03195.

(29) Shannon, R. D. Revised Effective Ionic Radii and Systematic Studies of Interatomic Distances in Halides and Chalcogenides. *Acta Crystallographica Section A* **1976**, *32* (5), 751–767. https://doi.org/10.1107/S0567739476001551.

(30) Kanai, S.; Heremans, F. J.; Seo, H.; Wolfowicz, G.; Anderson, C. P.; Sullivan, S. E.; Onizhuk, M.; Galli, G.; Awschalom, D. D.; Ohno, H. Generalized Scaling of Spin Qubit Coherence in over 12,000 Host Materials. *Proceedings of the National Academy of Sciences* **2022**, *119* (15), e2121808119. https://doi.org/10.1073/pnas.2121808119.

(31) Lea, K. R.; Leask, M. J. M.; Wolf, W. P. The Raising of Angular Momentum Degeneracy of F-Electron Terms by Cubic Crystal Fields. *Journal of Physics and Chemistry of Solids* **1962**, *23* (10), 1381–1405. https://doi.org/10.1016/0022-3697(62)90192-0.

(32) Ammerlaan, C. A. J.; de Maat-Gersdorf, I. Zeeman Splitting Factor of the Er3+ Ion in a Crystal Field. *Appl. Magn. Reson.* **2001**, *21* (1), 13–33. https://doi.org/10.1007/BF03162436.

(33) Calvin, J. J.; Kaufman, T. M.; Sedlak, A. B.; Crook, M. F.; Alivisatos, A. P. Observation of Ordered Organic Capping Ligands on Semiconducting Quantum Dots via Powder X-Ray Diffraction. *Nat Commun* **2021**, *12* (1), 2663. https://doi.org/10.1038/s41467-021-22947-x.

(34) Sims, C. M.; Maier, R. A.; Johnston-Peck, A. C.; Gorham, J. M.; Hackley, V. A.; Nelson, B. C. Approaches for the Quantitative Analysis of Oxidation State in Cerium Oxide Nanomaterials. *Nanotechnology* **2018**, *30* (8), 085703. https://doi.org/10.1088/1361-6528/aae364.

(35) Migani, A.; Vayssilov, G. N.; Bromley, S. T.; Illas, F.; Neyman, K. M. Dramatic Reduction of the Oxygen Vacancy Formation Energy in Ceria Particles: A Possible Key to Their Remarkable Reactivity at the Nanoscale. *J. Mater. Chem.* **2010**, *20* (46), 10535–10546. https://doi.org/10.1039/C0JM01908A.

(36) Yang, C.; Capdevila-Cortada, M.; Dong, C.; Zhou, Y.; Wang, J.; Yu, X.; Nefedov, A.; Heißler, S.; López, N.; Shen, W.; Wöll, C.; Wang, Y. Surface Refaceting Mechanism on Cubic Ceria. *J. Phys. Chem. Lett.* **2020**, *11* (18), 7925–7931. https://doi.org/10.1021/acs.jpclett.0c02409.

(37) Wu, L.; Wiesmann, H. J.; Moodenbaugh, A. R.; Klie, R. F.; Zhu, Y.; Welch, D. O.; Suenaga, M. Oxidation State and Lattice Expansion of CeO2-x Nanoparticles as a Function of Particle Size. *Phys. Rev. B* **2004**, *69* (12), 125415. https://doi.org/10.1103/PhysRevB.69.125415.

(38) Weber, M. J.; Bierig, R. W. Paramagnetic Resonance and Relaxation of Trivalent Rare-Earth Ions in Calcium Fluoride. I. Resonance Spectra and Crystal Fields. *Phys. Rev.* **1964**, *134* (6A), A1492–A1503. https://doi.org/10.1103/PhysRev.134.A1492.

(39) Grant, G. D.; Zhang, J.; Masiulionis, I.; Chattaraj, S.; Sautter, K. E.; Sullivan, S. E.; Chebrolu, R.; Liu, Y.; Martins, J. B.; Niklas, J.; Dibos, A. M.; Kewalramani, S.; Freeland, J. W.; Wen, J.; Poluektov, O. G.; Heremans, F. J.; Awschalom, D. D.; Guha, S. Optical and Microstructural Characterization of Er3+ Doped Epitaxial Cerium Oxide on Silicon. *APL Materials* **2024**, *12* (2), 021121. https://doi.org/10.1063/5.0181717.

(40) Johnston-Peck, A. C.; Winterstein, J. P.; Roberts, A. D.; DuChene, J. S.; Qian, K.; Sweeny, B. C.; Wei, W. D.; Sharma, R.; Stach, E. A.; Herzing, A. A. Oxidation-State Sensitive Imaging of Cerium Dioxide by Atomic-Resolution Low-Angle Annular Dark Field Scanning Transmission Electron Microscopy. *Ultramicroscopy* **2016**, *162*, 52–60. https://doi.org/10.1016/j.ultramic.2015.12.004.

(41) Johnston-Peck, A. C.; Yang, W.-C. D.; Winterstein, J. P.; Sharma, R.; Herzing, A. A. In Situ Oxidation and Reduction of Cerium Dioxide Nanoparticles Studied by Scanning Transmission Electron Microscopy. *Micron* **2018**, *115*, 54–63. https://doi.org/10.1016/j.micron.2018.08.008.

(42) Bauch, E.; Singh, S.; Lee, J.; Hart, C. A.; Schloss, J. M.; Turner, M. J.; Barry, J. F.; Pham, L. M.; Bar-Gill, N.; Yelin, S. F.; Walsworth, R. L. Decoherence of Ensembles of Nitrogen-Vacancy Centers in Diamond. *Phys. Rev. B* **2020**, *102* (13), 134210. https://doi.org/10.1103/PhysRevB.102.134210.

(43) Zhang, J.; Grant, G. D.; Masiulionis, I.; Solomon, M. T.; Bindra, J. K.; Niklas, J.; Dibos, A. M.; Poluektov, O. G.; Heremans, F. J.; Guha, S.; Awschalom, D. D. Optical and Spin Coherence of Er3+ in Epitaxial CeO2 on Silicon. arXiv September 28, 2023. https://doi.org/10.48550/arXiv.2309.16785.

(44) Stapleton, H. J.; Brower, K. L. Electron-Spin-Resonance Linewidth Variation with Temperature in Some Rare-Earth Salts: T1/T2 Ratios. *Phys. Rev.* **1969**, *178* (2), 481–485. https://doi.org/10.1103/PhysRev.178.481.

(45) Orbach, R. Spin-Lattice Relaxation in Rare-Earth Salts: Field Dependence of the Two-Phonon Process. *Proceedings of the Royal Society of London. Series A. Mathematical and Physical Sciences* **1961**, *264* (1319), 485–495. https://doi.org/10.1098/rspa.1961.0212.

(46) Jeremy Amdur, M.; R. Mullin, K.; J. Waters, M.; Puggioni, D.; K. Wojnar, M.; Gu, M.; Sun, L.; H. Oyala, P.; M. Rondinelli, J.; E. Freedman, D. Chemical Control of Spin–Lattice Relaxation to Discover a Room Temperature Molecular Qubit. *Chemical Science* **2022**, *13* (23), 7034–7045. https://doi.org/10.1039/D1SC06130E.

(47) Guo, X.; Stramma, A. M.; Li, Z.; Roth, W. G.; Huang, B.; Jin, Y.; Parker, R. A.; Arjona Martínez, J.; Shofer, N.; Michaels, C. P.; Purser, C. P.; Appel, M. H.; Alexeev, E. M.; Liu, T.; Ferrari, A. C.; Awschalom, D. D.; Delegan, N.; Pingault, B.; Galli, G.; Heremans, F. J.; Atatüre, M.; High, A. A. Microwave-Based Quantum Control and Coherence Protection of Tin-Vacancy Spin Qubits in a Strain-Tuned Diamond-Membrane Heterostructure. *Phys. Rev. X* **2023**, *13* (4), 041037. https://doi.org/10.1103/PhysRevX.13.041037.

(48) Zhong, M.; Hedges, M. P.; Ahlefeldt, R. L.; Bartholomew, J. G.; Beavan, S. E.; Wittig, S. M.; Longdell, J. J.; Sellars, M. J. Optically Addressable Nuclear Spins in a Solid with a Six-Hour Coherence Time. *Nature* **2015**, *517* (7533), 177–180. https://doi.org/10.1038/nature14025.

(49) Yang, J.; Fan, W.; Zhang, Y.; Duan, C.; de Boo, G. G.; Ahlefeldt, R. L.; Longdell, J. J.; Johnson, B. C.; McCallum, J. C.; Sellars, M. J.; Rogge, S.; Yin, C.; Du, J. Zeeman and Hyperfine Interactions of a Single $^{167}Er^{3+}$ Ion in Si. *Phys. Rev. B* **2022**, *105* (23), 235306. https://doi.org/10.1103/PhysRevB.105.235306.

(50) Longdell, J. J.; Alexander, A. L.; Sellars, M. J. Characterization of the Hyperfine Interaction in Europium-Doped Yttrium Orthosilicate and Europium Chloride Hexahydrate. *Phys. Rev. B* **2006**, *74* (19), 195101. https://doi.org/10.1103/PhysRevB.74.195101.

(51) Ohno, K.; Joseph Heremans, F.; Bassett, L. C.; Myers, B. A.; Toyli, D. M.; Bleszynski Jayich, A. C.; Palmstrøm, C. J.; Awschalom, D. D. Engineering Shallow Spins in Diamond with Nitrogen Delta-Doping. *Applied Physics Letters* **2012**, *101* (8), 082413. https://doi.org/10.1063/1.4748280.

(52) Gupta, J. A.; Awschalom, D. D.; Efros, Al. L.; Rodina, A. V. Spin Dynamics in Semiconductor Nanocrystals. *Phys. Rev. B* **2002**, *66* (12), 125307. https://doi.org/10.1103/PhysRevB.66.125307.

(53) Bloembergen, N.; Shapiro, S.; Pershan, P. S.; Artman, J. O. Cross-Relaxation in Spin Systems. *Phys. Rev.* **1959**, *114* (2), 445–459. https://doi.org/10.1103/PhysRev.114.445.

(54) Yang, W.; Liu, R.-B. Quantum Many-Body Theory of Qubit Decoherence in a Finite-Size Spin Bath. *Phys. Rev. B* **2008**, *78* (8), 085315. https://doi.org/10.1103/PhysRevB.78.085315.

(55) Peng, X.; Manna, L.; Yang, W.; Wickham, J.; Scher, E.; Kadavanich, A.; Alivisatos, A. P. Shape Control of CdSe Nanocrystals. *Nature* **2000**, *404* (6773), 59–61. https://doi.org/10.1038/35003535.

(56) Agnello, S.; Boscaino, R.; Cannas, M.; Gelardi, F. M. Instantaneous Diffusion Effect on Spin-Echo Decay: Experimental Investigation by Spectral Selective Excitation. *Phys. Rev. B* **2001**, *64* (17), 174423. https://doi.org/10.1103/PhysRevB.64.174423.

(57) Welinski, S.; Woodburn, P. J. T.; Lauk, N.; Cone, R. L.; Simon, C.; Goldner, P.; Thiel, C. W. Electron Spin Coherence in Optically Excited States of Rare-Earth Ions for Microwave to Optical Quantum Transducers. *Phys. Rev. Lett.* **2019**, *122* (24), 247401. https://doi.org/10.1103/PhysRevLett.122.247401.

(58) Bryan, J. D.; Schwartz, D. A.; Gamelin, D. R. The Influence of Dopants on the Nucleation of Semiconductor Nanocrystals from Homogeneous Solution. *Journal of Nanoscience and Nanotechnology* **2005**, *5* (9), 1472–1479. https://doi.org/10.1166/jnn.2005.314.

(59) Tyryshkin, A. M.; Tojo, S.; Morton, J. J. L.; Riemann, H.; Abrosimov, N. V.; Becker, P.; Pohl, H.-J.; Schenkel, T.; Thewalt, M. L. W.; Itoh, K. M.; Lyon, S. A. Electron Spin Coherence Exceeding Seconds in High-Purity Silicon. *Nature Mater* **2012**, *11* (2), 143–147. https://doi.org/10.1038/nmat3182.

(60) Witzel, W. M.; Carroll, M. S.; Morello, A.; Cywiński, Ł.; Das Sarma, S. Electron Spin Decoherence in Isotope-Enriched Silicon. *Phys. Rev. Lett.* **2010**, *105* (18), 187602. https://doi.org/10.1103/PhysRevLett.105.187602.

(61) Mims, W. B. Phase Memory in Electron Spin Echoes, Lattice Relaxation Effects in CaW${\mathrm{O}}_{4}$: Er, Ce, Mn. *Phys. Rev.* **1968**, *168* (2), 370–389. https://doi.org/10.1103/PhysRev.168.370.

(62) Schnitzer, I.; Yablonovitch, E.; Caneau, C.; Gmitter, T. J. Ultrahigh Spontaneous Emission Quantum Efficiency, 99.7% Internally and 72% Externally, from AlGaAs/GaAs/AlGaAs Double Heterostructures. *Applied Physics Letters* **1993**, *62* (2), 131–133. https://doi.org/10.1063/1.109348.

(63) Othman, A.; Gowda, A.; Andreescu, D.; Hassan, M. H.; Babu, S. V.; Seo, J.; Andreescu, S. Two Decades of Ceria Nanoparticle Research: Structure, Properties and Emerging Applications. *Mater. Horiz.* **2024**. https://doi.org/10.1039/D4MH00055B.

(64) Paier, J.; Penschke, C.; Sauer, J. Oxygen Defects and Surface Chemistry of Ceria: Quantum Chemical Studies Compared to Experiment. *Chem. Rev.* **2013**, *113* (6), 3949–3985. https://doi.org/10.1021/cr3004949.

(65) Boles, M. A.; Ling, D.; Hyeon, T.; Talapin, D. V. The Surface Science of Nanocrystals. *Nature Mater* **2016**, *15* (2), 141–153. https://doi.org/10.1038/nmat4526.

(66) Owen, J. The Coordination Chemistry of Nanocrystal Surfaces. *Science* **2015**, *347* (6222), 615–616. https://doi.org/10.1126/science.1259924.

(67) Dwyer, B. L.; Rodgers, L. V. H.; Urbach, E. K.; Bluvstein, D.; Sangtawesin, S.; Zhou, H.; Nassab, Y.; Fitzpatrick, M.; Yuan, Z.; De Greve, K.; Peterson, E. L.; Knowles, H.; Sumarac, T.; Chou, J.-P.; Gali, A.; Dobrovitski, V. V.; Lukin, M. D.; de Leon, N. P. Probing Spin Dynamics on Diamond Surfaces Using a Single Quantum Sensor. *PRX Quantum* **2022**, *3* (4), 040328. https://doi.org/10.1103/PRXQuantum.3.040328.

(68) Choi, J.; Zhou, H.; Knowles, H. S.; Landig, R.; Choi, S.; Lukin, M. D. Robust Dynamic Hamiltonian Engineering of Many-Body Spin Systems. *Phys. Rev. X* **2020**, *10* (3), 031002. https://doi.org/10.1103/PhysRevX.10.031002.

(69) Álvarez, G. A.; Suter, D. Measuring the Spectrum of Colored Noise by Dynamical Decoupling. *Phys. Rev. Lett.* **2011**, *107* (23), 230501. https://doi.org/10.1103/PhysRevLett.107.230501.

(70) Ourari, S.; Dusanowski, Ł.; Horvath, S. P.; Uysal, M. T.; Phenicie, C. M.; Stevenson, P.; Raha, M.; Chen, S.; Cava, R. J.; de Leon, N. P.; Thompson, J. D. Indistinguishable Telecom Band Photons from a Single Er Ion in the Solid State. *Nature* **2023**, *620* (7976), 977–981. https://doi.org/10.1038/s41586-023-06281-4.

(71) Yang, S.; Gao, L. Controlled Synthesis and Self-Assembly of CeO2 Nanocubes. *J. Am. Chem. Soc.* **2006**, *128* (29), 9330–9331. https://doi.org/10.1021/ja063359h.

(72) Morrison, C.; Sun, H.; Yao, Y.; Loomis, R. A.; Buhro, W. E. Methods for the ICP-OES Analysis of Semiconductor Materials. *Chem. Mater.* **2020**, *32* (5), 1760–1768. https://doi.org/10.1021/acs.chemmater.0c00255.

(73) Wang, X.; Li, J.; Ha, H. D.; Dahl, J. C.; Ondry, J. C.; Moreno-Hernandez, I.; Head-Gordon, T.; Alivisatos, A. P. AutoDetect-mNP: An Unsupervised Machine Learning Algorithm for Automated Analysis of Transmission Electron Microscope Images of Metal Nanoparticles. *JACS Au* **2021**, *1* (3), 316–327. https://doi.org/10.1021/jacsau.0c00030.

(74) Stoll, S.; Schweiger, A. EasySpin, a Comprehensive Software Package for Spectral Simulation and Analysis in EPR. *Journal of Magnetic Resonance* **2006**, *178* (1), 42–55. https://doi.org/10.1016/j.jmr.2005.08.013.

(75) Slusher, R. E.; Hahn, E. L. Sensitive Detection of Nuclear Quadrupole Interactions in Solids. *Phys. Rev.* **1968**, *166* (2), 332–347. https://doi.org/10.1103/PhysRev.166.332.

(76) Olivero, J. J.; Longbothum, R. L. Empirical Fits to the Voigt Line Width: A Brief Review. *Journal of Quantitative Spectroscopy and Radiative Transfer* **1977**, *17* (2), 233–236. https://doi.org/10.1016/0022-4073(77)90161-3.

(77) Onizhuk, M.; Galli, G. PyCCE: A Python Package for Cluster Correlation Expansion Simulations of Spin Qubit Dynamics. *Advanced Theory and Simulations* **2021**, *4* (11), 2100254. https://doi.org/10.1002/adts.202100254.

(78)  Onizhuk, M.; Wang, Y.-X.; Nagura, J.; Clerk, A. A.; Galli, G. Understanding Central Spin Decoherence Due to Interacting Dissipative Spin Baths. arXiv December 2, 2023. https://doi.org/10.48550/arXiv.2312.01205.

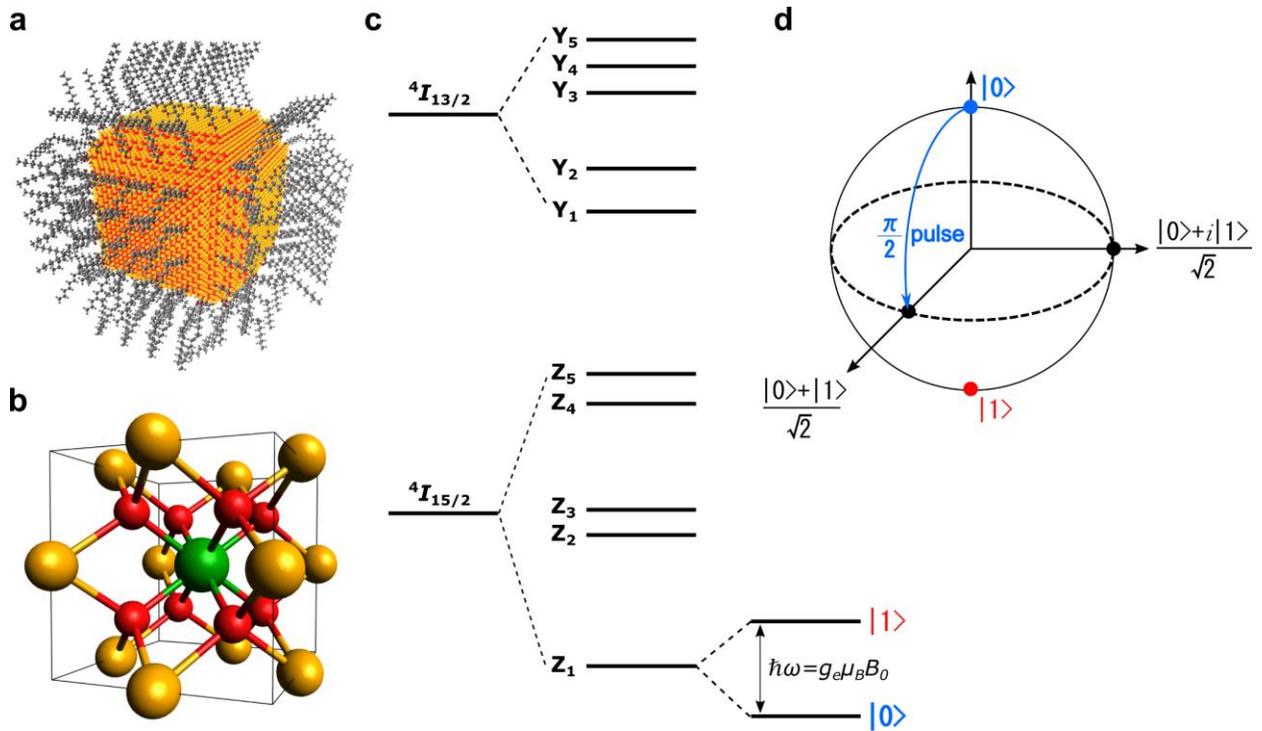

Figure 1. Overview of $Er^{3+}$ doped $CeO_2$ nanocrystals. (a) Schematic of a nanocrystal with an inorganic core (cerium dioxide doped with erbium) and an outer shell of organic (oleate) ligands. (b) A unit cell of the cerium dioxide with a substitutional erbium defect (green) is depicted. Red atoms are oxygen while orange atoms are cerium. (c) Ground state ($^4I_{15/2}$) and first excited state ($^4I_{13/2}$) of $Er^{3+}$ in a cubic crystal field environment. The crystal field splitting in a cubic symmetry split both the ground states ($Z_n$) and excited states ($Y_n$) into five energy levels. Also depicted is the Zeeman splitting of the ground state, which is used as a spin qubit. (d) Bloch sphere representation of the spin qubit. Microwave pulses are used to rotate the spin qubit along the Bloch sphere.

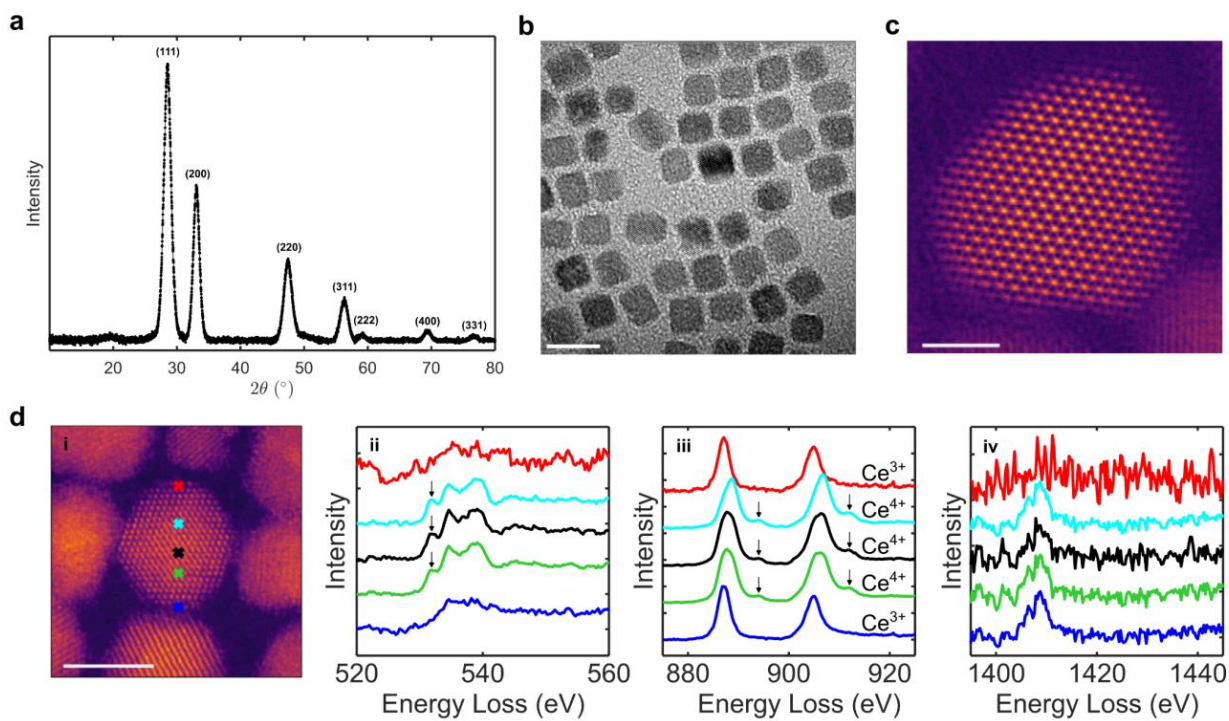

Figure 2. Structure and chemical composition of $Er^{3+}$ doped $CeO_2$ nanocrystals. (a) Typical powder x-ray diffractogram of the erbium doped ceria nanocrystals with different crystal diffraction planes labeled. (b) Bright field transmission electron micrograph of erbium doped ceria nanocrystals. The scale bar is 10 nm. (c) Inverted ABF image of an erbium doped ceria nanocrystal oriented along the (110) facet, depicting a highly crystalline structure. The scale bar is 2 nm. (d) Spatially resolved EEL spectra collected at different probe positions labeled in the HAADF image in (i) of an erbium doped ceria nanocrystal ($\langle N_{Er} \rangle = 179.8$). The scale bar is 5 nm. The extracted EEL spectra for O K, Ce M and Er M edges are shown in (ii), (iii) and (iv) respectively. The arrows in (ii) and (iii) mark the O K edge pre-peak and Ce M post-edge peaks respectively.

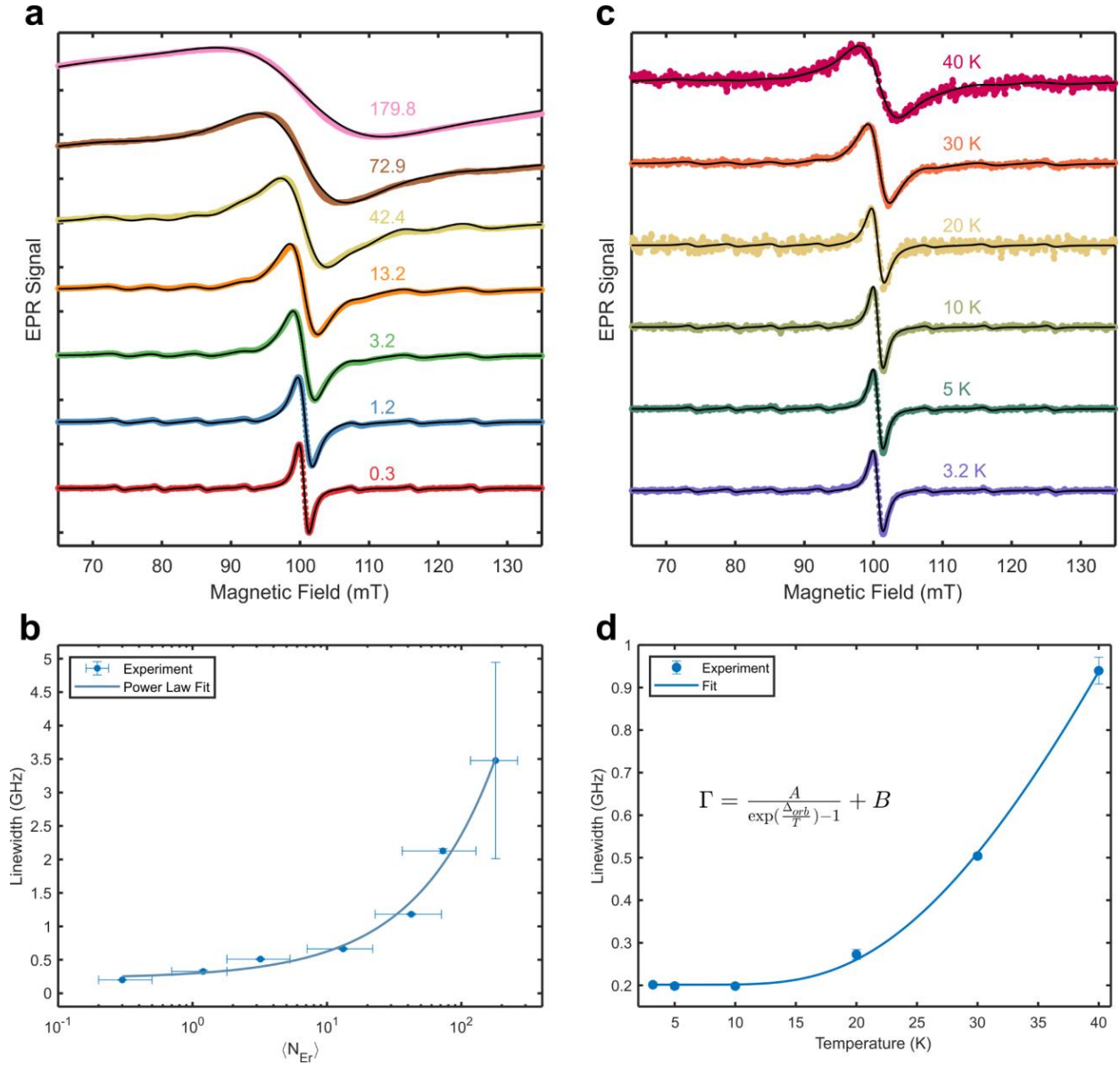

Figure 3. Concentration and temperature dependence of EPR linewidth. (a) Experimental (colored dots) and fitted (black solid lines) CW-EPR spectra of erbium doped ceria nanocrystals with different average number of erbium dopants per nanocrystal denoted nearby each spectrum. Measurement temperatures were ~3.2 K. (b) Fitted linewidth for different concentrations of erbium per nanocrystal ($I = 0$). The fitted power law exponent is 0.73±0.26. Horizontal error bars come from the standard deviation of the size distribution and vertical error bars are 95% confidence intervals for the linewidth fit, smaller than the size of the marker for many data points. (c) Experimental CW-EPR spectra as a function of temperature for $\langle N_{Er} \rangle = 0.3$. Each experimental spectrum (colored dots) is fitted to a simulated spectrum (black solid lines) with linewidth broadening as a fitting parameter. (d) Fitted linewidth for different experimental temperatures with error bars corresponding to a 95% confidence interval. The linewidth dependence is fitted to an Orbach term along with a constant offset which accounts for inhomogeneous broadening. Fitted values are $A = 7.7 \pm 2.7$ GHz, $\Delta_{orb} = 98 \pm 13$ K, $B = 0.20 \pm 0.02$ GHz.

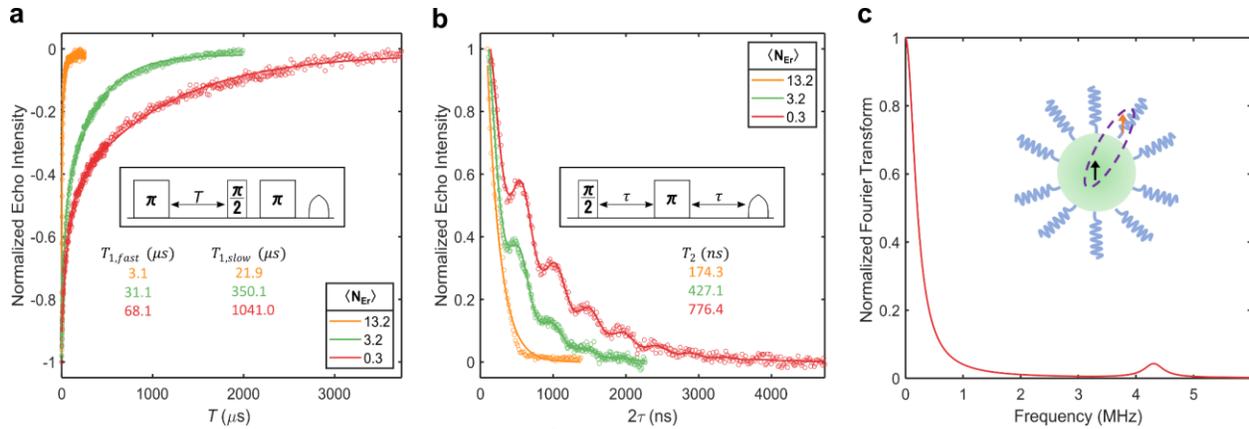

Figure 4. Spin relaxation and spin coherence of $Er^{3+}$ doped $CeO_2$ nanocrystals. (a) Echo-detected inversion recovery for different concentrations of erbium dopants per nanocrystal, corresponding to the spin-lattice relaxation time $T_1$ of the electron spin. Biexponential fits are shown as solid lines, with corresponding fitted values listed. Inset corresponds to the pulse sequence for measuring inversion recovery. (b) Electron coherence measured using a Hahn-echo sequence for different concentrations of erbium dopants per nanocrystal, corresponding to the $T_2$ of the electron spin. Solid lines correspond to a Fourier transformed Lorentzian model fit, with extracted spin-spin relaxation time $T_2$ values listed. Inset corresponds to the Hahn-echo pulse sequence for measuring spin coherence. Measurement temperatures for (a) and (b) were ~3.2 K. (c) Fourier transform of the fitted time trace of the $\langle N_{Er} \rangle = 0.3$ signal, inset corresponds to a schematic depicting the coupling of erbium ions to hydrogen nuclei on the surface.

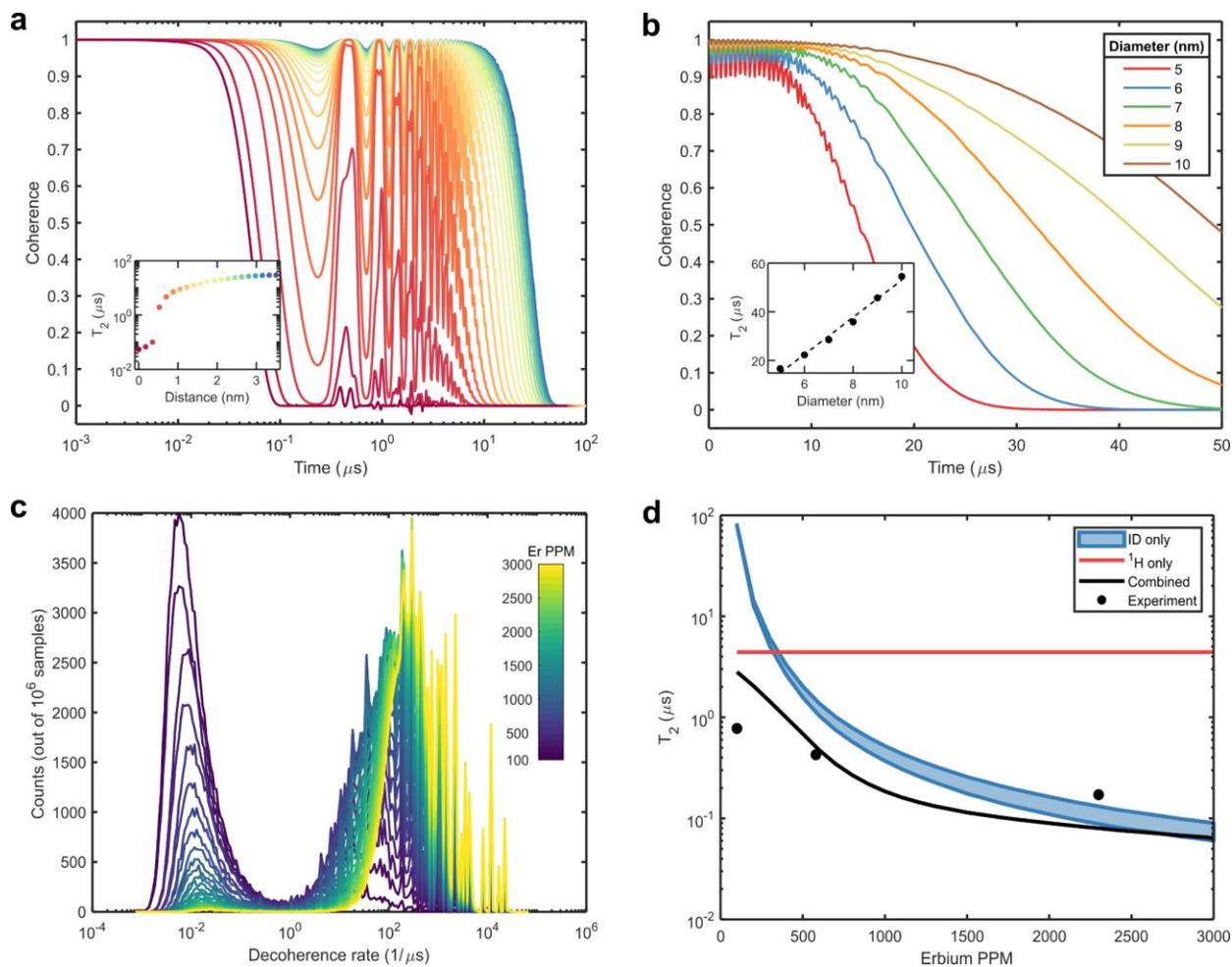

Figure 5. CCE simulations of $Er^{3+}$ doped $CeO_2$ nanocrystals. (a) Calculated spin coherence decay as a function of radial position from the surface (0 nm) to the center (3.5 nm) of a 7 nm diameter $CeO_2$ nanocrystal. Inset corresponds to the extracted $T_2$ decoherence time, with each colored dot corresponding to the respectively colored curve. (b) Calculated spin coherence decay as a function of the diameter of the $CeO_2$ nanocrystal, for an erbium spin defect located at the center of the nanocrystal. Inset corresponds to the extracted $T_2$ decoherence time. (c) Calculated distribution of decoherence rates due solely to instantaneous diffusion of neighboring erbium spins. Colors correspond to different erbium molar ppm (d). Calculated decoherence due to instantaneous diffusion (blue), due to hydrogen nuclei (red) and combining both effects (black). Also plotted are the experimentally measured $T_2$ values.